\documentclass[12pt, fleqn]{iopart}

\usepackage{iopams}

\usepackage{graphicx}
\usepackage{theorem}
\usepackage{hyperref}

\theoremstyle{plain}
\newtheorem{Thm}{Theorem}
\newtheorem{Lemma}[Thm]{Lemma}

\newtheorem{Def}[Thm]{Definition}

\theorembodyfont{\rmfamily}
\newtheorem{Proof}{Proof}

\begin{document}

\title{Solutions to the ultradiscrete KP hierarchy and its reductions}
\author{Yoichi Nakata}
\address{$^1$ Interdisciplinary Center for Mathematical Sciences, Graduate School of Mathematical Sciences, the University of Tokyo, 3-8-1 Komaba, Meguro-ku, 153-8914 Tokyo, Japan}
\address{$^2$ Institute for Biology and Mathematics of Dynamical Cell Processes (iBMath), the University of Tokyo, 3-8-1 Komaba, Meguro-ku, 153-8914 Tokyo, Japan}
\ead{ynakata@ms.u-tokyo.ac.jp}
\begin{abstract}
We propose a recursive representation of solutions to an ultradiscrete analogue of the discrete KP hierarchy, which is the master equation of discrete soliton equations. We also propose a class of solutions which can be used to start the recursion. Finally, as an application, we discuss the compatibility condition of ultradiscrete soliton equations.
\end{abstract}
\vspace{2pc}

\section{Introduction}

In this paper, we consider an equation named after two important topics in the field of integrable systems---the KP hierarchy and the ultradiscrete systems.

The KP hierarchy \cite{DataJimboKashiwaraMiwa} is a sequence of equations generated from a bilinear identity and plays an important role in the construction of a unified theory for integrable systems. By applying a transformation for the independent variables called the Miwa transformation, the KP hierarchy is rewritten in the form of the discrete KP hierarchy \cite{OhtaHirotaTsujimotoImai}. The fundamental equation in the discrete KP hierarchy is a trinomial equation known as the Hirota-Miwa equation \cite{Hirota1981, Miwa}. It is known that all equations in the discrete KP hierarchy are generated by such trinomial equations \cite{PrivateComm1} and that most discrete integrable equations are obtained by taking some reduction for this equation.

Ultradiscrete systems are difference equations in which only $\max$ and $\pm$ operators appear and which are obtained by a limiting procedure \cite{TokihiroTakahashiMatsukidairaSatsuma} from canonical difference equations. The remarkable point of this procedure is that it preserves the good properties of integrable systems, although the dependent variables only take discrete values. The most famous example is the Box and Ball system (BBS) \cite{TakahashiSatsuma}, which is a cellular automaton consisting of an infinite sequence of boxes and a finite amount of balls. The BBS has solitons and an infinite amount of conserved quantities and is obtained by the ultradiscretization of the KdV equation.

It is an interesting problem to try to obtain the structure of the solutions to ultradiscrete soliton equations, as opposed to those of ordinary soliton equations.
In previous papers \cite{Nakata2009, Nakata2010}, we proposed a recursive representation which corresponds to the notion of vertex operators, to several ultradiscrete analogues of soliton equations including the ultradiscrete KP equation. As an analogue of determinant-type solutions, the ultradiscretization of signature-free determinants is discussed in \cite{TakahashiHirota} and the relationship between this type of solution and ultradiscrete soliton equations is discussed in \cite{NagaiTakahashi2010, NagaiTakahashi2011}. The establishment of relationships to other mathematical topics is also studied, for example, algebro-geometrical \cite{KimijimaTokihiro, InoueTakenawa, Iwao2009} and combinatorial theories \cite{TakagakiKamioka, NoumiYamada, Nakata2011}.

Recently, finding exact solutions to specific ultradiscrete systems became an active field of research. For example, negative soliton solutions were found in the analysis of the BBS with generalized dependent variables \cite{Hirota2009}.  We previously solved the initial value problem of this equation including negative solitons \cite{WilloxNakataSatsumaRamaniGrammaticos}. Soliton solutions with periodic phase term are presented in \cite{Nakamura}. Such solutions do not appear in discrete integrable equations.

In this paper, we first propose yet another solution of the ultradiscrete KP equation. We secondly extend this solution to that of the ultradiscrete KP hierarchy and the non-autonomous one. We next discuss ultradiscrete specific solutions called backgrounds to this equation. Finally, as an application, we consider the background solutions of the BBS with Carrier (BBSC), express this cellular automata as a Lax form and discuss its compatibility condition. It is expected that the ultradiscrete KP hierarchy can be the master equation of the ultradiscrete as well as discrete systems.

\section{An extension of solutions to the ultradiscrete KP equation}

\begin{Def}
We define $T^{(N)}_{n,n',k,l,m}$ by
\begin{eqnarray}\label{defkphtau}
   T^{(N)}_{n,n',k,l,m}  = \max \Big( T^{(N-1)}_{n,n',k,l,m}, \eta_N + T^{(N-1)}_{n-1,n'-1,k,l,m} \Big) \qquad ( N \ge 1 ) \\
   T^{(0)}_{n,n',k,l,m} \equiv 0
\end{eqnarray}
for $\eta_N$ given by $\eta_N = n P_N + n' Q_N + k \Xi_N + l \Omega_N + m \Omega'_N + C_N$. Here, $P_N$ and $Q_N$ are parameters satisfying
\begin{eqnarray}
  0 < P_1 \le P_2 \le \ldots \le P_N \\
  0 < Q_1 \le Q_2 \le \ldots \le Q_N,
\end{eqnarray}
$\Xi_N$, $\Omega_N$ and $\Omega'_N$ are given by
\begin{eqnarray}\label{defomega}
  \Xi_N = \min ( S, P_N ) \\
  \Omega_N = \min ( R, Q_N ) \\
  \Omega'_N = \min ( R', Q_N ),
\end{eqnarray}
where $S>0$, $R \ge R' > 0$.
\end{Def}

\begin{Thm}\label{mainThm}
$T^{(N)}_{n,n',k,l,m}$ satisfies equations:
\begin{eqnarray}
  T^{(N)}_{n, n', k, l, m+1} + T^{(N)}_{n, n', k+1, l+1, m} \nonumber\\
  = \max \Big( T^{(N)}_{n, n', k+1, l+1, m+1} + T^{(N)}_{n, n', k, l, m} - R', T^{(N)}_{n, n', k, l+1, m} + T^{(N)}_{n, n', k+1, l, m+1} \Big) \label{ukphier1} \\
  T^{(N)}_{n,n',k,l,m+1} + T^{(N)}_{n-1,n'-1,k+1,l+1,m} = \max \Big( T^{(N)}_{n-1,n'-1,k+1,l+1,m+1} + T^{(N)}_{n,n',k,l,m}, \nonumber\\
  T^{(N)}_{n,n',k,l+1,m} + T^{(N)}_{n-1,n'-1,k+1,l,m+1} - R + R', \nonumber\\ 
  T^{(N)}_{n-1,n'-1,k,l+1,m} + T^{(N)}_{n,n',k+1,l,m+1} - S \Big). \label{ukphier2}
\end{eqnarray}
\end{Thm}

The first equation (\ref{ukphier1}) does not depend on $n$ and $n'$. Then, by regarding $n$ and $n'$ as parameters, this equation is the ultradiscrete KP equation we proposed in \cite{Nakata2010}. We shall call this equation the ultradiscrete KP equation for the variables ($k; l, m$). By focusing on variables $n$, $n'$ and $m$ and regarding the other variables as constants, the function (\ref{defkphtau}) is equal to the solutions we proposed previously in \cite{Nakata2010}. Therefore, it solves the equation
\begin{eqnarray}\label{ukpnnpm}
  T^{(N)}_{n, n', k, l, m+1} + T^{(N)}_{n+1, n'+1, k, l, m} \nonumber\\
  = \max \Big( T^{(N)}_{n+1, n'+1, k, l, m+1} + T^{(N)}_{n, n', k, l, m} - R', T^{(N)}_{n, n'+1, k, l, m} + T^{(N)}_{n+1, n', k, l, m+1} \Big).
\end{eqnarray}
We call this equation the ultradiscrete KP equation for the variables ($n$; $n'$, $m$). This corresponds to (\ref{ukphier1}) in the case where $S$ and $R$ are larger than $P_N$ and $Q_N$ respectively and the function depends on the form of $n+k$ and $n'+l$ for independent variables $n$, $n'$, $k$ and $l$.  We note that this function also solves
\begin{eqnarray}\label{uknpnk}
  T^{(N)}_{n, n', k+1, l, m} + T^{(N)}_{n+1, n'+1, k, l, m} \nonumber\\
  = \max \Big( T^{(N)}_{n+1, n'+1, k+1, l, m} + T^{(N)}_{n, n', k, l, m} - S, T^{(N)}_{n+1, n', k, l, m} + T^{(N)}_{n, n'+1, k+1, l, m} \Big).
\end{eqnarray}
by interchanging the roles of ($n$, $k$) and ($n'$, $m$). We call this equation the ultradiscrete KP equation for the variables ($n'; n, k$). We also note that if the function (\ref{defkphtau}) solves (\ref{ukphier1}), it also solves
\begin{eqnarray}\label{ukpmnk}
  T^{(N)}_{n, n', k+1, l, m} + T^{(N)}_{n+1, n', k, l, m+1} \nonumber\\
  = \max \Big( T^{(N)}_{n+1, n', k+1, l, m+1} + T^{(N)}_{n, n', k, l, m} - S, T^{(N)}_{n+1, n', k, l, m} + T^{(N)}_{n, n', k+1, l, m+1} \Big).
\end{eqnarray}
by combining arguments above. We call this equation the ultradiscrete KP equation for variables the ($m; n, k$).

Before proceeding with the proof, we introduce some lemmas.

\begin{Lemma}\label{Lem1}
Let 
\begin{eqnarray}
  H^{(N)}_{n,n',k,l,m} = T^{(N)}_{n+\tilde{n},n'+\tilde{n}',k+\tilde{k},l+\tilde{l},m+\tilde{m}} + T^{(N)}_{n-1,n'-1,k,l,m} \nonumber\\
	\qquad\qquad\qquad\qquad  - T^{(N)}_{n+\tilde{n}-1,n'+\tilde{n}'-1,k+\tilde{k},l+\tilde{l},m+\tilde{m}} - T^{(N)}_{n,n',k,l,m}
\end{eqnarray}
for $\tilde{n}, \tilde{n}', \tilde{k}, \tilde{l}, \tilde{m}$ such that
\begin{equation}
	0 \le \tilde{n} P_1 + \tilde{n}' Q_1 + \tilde{k} \Xi_1 + \tilde{l} \Omega_1 + \tilde{m} \Omega'_1 \le \ldots \le \tilde{n} P_N + \tilde{n}' Q_N + \tilde{k} \Xi_N + \tilde{l} \Omega_N + \tilde{m} \Omega'_N.
\end{equation}
Then, it holds that
\begin{equation}
  H^{(N)}_{n,n',k,l,m} \le \tilde{n} P_N + \tilde{n}' Q_N + \tilde{k} \Xi_N + \tilde{l} \Omega_N + \tilde{m} \Omega'_N.
\end{equation}
\end{Lemma}
We omit the proof because it is the same as that of Lemma 2 in \cite{Nakata2010}.

\begin{Lemma}\label{Lem2}
When one requires that $T^{(N)}_{n,n',l,m,n}$ are solutions of (\ref{ukphier1}) and (\ref{ukphier2}), one then has
\begin{eqnarray}
	T^{(N)}_{n,n',k,l+1,m} + T^{(N)}_{n-1,n'-1,k+1,l,m+1} - T^{(N)}_{n,n',k,l,m+1} - T^{(N)}_{n-1,n'-1,k+1,l+1,m} \nonumber\\ 
	\le \Omega_N - \Omega'_N \label{Lem2-1} \\
	T^{(N)}_{n-1,n'-1,k,l+1,m} + T^{(N)}_{n,n',k+1,l,m+1} - T^{(N)}_{n,n',k,l,m+1} - T^{(N)}_{n-1,n'-1,k+1,l+1,m} \le \Xi_N \label{Lem2-2} \\
	T^{(N)}_{n-2,n'-2,k+1,l+1,m+1} + T^{(N)}_{n,n',k,l,m} - T^{(N)}_{n-1,n'-1,k,l,m+1} - T^{(N)}_{n-1,n'-1,k+1,l+1,m} \nonumber\\
		 \le P_N - \Xi_N + Q_N - \Omega_N \label{Lem2-3}
\end{eqnarray}
for $N \ge 1$.
\end{Lemma}

\begin{Proof}
We prove (\ref{Lem2-1}) by induction. By employing $\max(a, b)  - \max(c, d) \le \max(a-b, c-d)$, we obtain
\begin{eqnarray}
	T^{(N)}_{n,n',k,l+1,m} - T^{(N)}_{n,n',k,l,m+1} \le \max ( T^{(N-1)}_{n,n',k,l+1,m} - T^{(N-1)}_{n,n',k,l,m+1}, \nonumber\\
								       \Omega_N - \Omega'_N + T^{(N-1)}_{n-1,n'-1,k,l+1,m} - T^{(N-1)}_{n-1,n'-1,k,l,m+1} ) \\
	T^{(N)}_{n-1,n'-1,k+1,l,m+1} - T^{(N)}_{n-1,n'-1,k+1,l+1,m} \le \max ( T^{(N-1)}_{n-1,n'-1,k+1,l,m+1} - T^{(N-1)}_{n-1,n'-1,k+1,l+1,m}, \nonumber\\
								    -(\Omega_N - \Omega'_N) + T^{(N-1)}_{n-2,n'-2,k+1,l,m+1} - T^{(N-1)}_{n-2,n'-2,k+1,l+1,m} ).
\end{eqnarray}
Adding the inequalities yields
\begin{eqnarray}
	H'^{(N)}_{n,n',k,l,m} = \max \Big( H'^{(N-1)}_{n,n',k,l,m}, H'^{(N-1)}_{n,n',k,l,m}, \nonumber\\
				           \Omega_N - \Omega'_N + T^{(N-1)}_{n-1,n'-1,k,l+1,m} + T^{(N-1)}_{n-1,n'-1,k+1,l,m+1} \nonumber\\ - T^{(N-1)}_{n-1,n'-1,k,l,m+1} - T^{(N-1)}_{n-1,n'-1,k+1,l+1,m}, \nonumber\\
					-(\Omega_N - \Omega'_N) + H'^{(N-1)}_{n,n',k,l,m}  + H^{(N-1)}_{n,n',k,l,m}\Big|_{\tilde{n}=0, \tilde{n'}=0, \tilde{k}=0, \tilde{l}=1, \tilde{m}=-1} \Big),
\end{eqnarray}
where $H^{(N)}_{n,n',k,l,m} = T^{(N)}_{n,n',k,l+1,m} + T^{(N)}_{n-1,n'-1,k+1,l,m+1} - T^{(N)}_{n,n',k,l,m+1} - T^{(N)}_{n-1,n'-1,k+1,l+1,m}$. The third argument is less than $\Omega_N - \Omega'_N$ because of (\ref{ukphier1}) for $N-1$. Then, all arguments in the maximum are less than $\Omega_N - \Omega'_N$ by the assumption of the induction and Lemma \ref{Lem1}.

The proofs of (\ref{Lem2-2}) and (\ref{Lem2-3}) are the same. 
\end{Proof}

Now, let us prove Theorem \ref{mainThm}.

\begin{Proof}
We employ induction on N. It is trivial that $T^{(0)}_{n, n', k, l, m} \equiv 0$ solves these equations and we assume that $T^{(N-1)}_{n,n',k,l,m}$ satisfies (\ref{ukphier1}) and (\ref{ukphier2}). Let us prove (\ref{ukphier1}) first. By definition (\ref{defkphtau}), one has
\begin{eqnarray}
  T^{(N)}_{n,n',k,l,m+1} + T^{(N)}_{n,n',k+1,l+1,m} = \max \Big( T^{(N-1)}_{n,n',k,l,m+1} + T^{(N-1)}_{n,n',k+1,l+1,m}, \nonumber\\ 
 \Xi_N + \Omega_N + \eta_{N} + T^{(N-1)}_{n,n',k,l,m+1} + T^{(N-1)}_{n-1,n'-1,k+1,l+1,m}, \nonumber\\ 
 \Omega'_N + \eta_{N} + T^{(N-1)}_{n-1,n'-1,k,l,m+1} + T^{(N-1)}_{n,n',k+1,l+1,m}, \nonumber\\
 \Xi_N + \Omega_N + \Omega'_N + 2 \eta_{N} + T^{(N-1)}_{n-1,n'-1,k,l,m+1} + T^{(N-1)}_{n-1,n'-1,k+1,l+1,m} \Big). \label{ukph1}
\end{eqnarray}
By virtue of Lemma \ref{Lem1}, the third argument of (\ref{ukph1}) cannot yield the maximum because it is always less than the second. Therefore, we can rewrite (\ref{ukph1}) as
\begin{eqnarray}
  T^{(N)}_{n,n',k,l,m+1} + T^{(N)}_{n,n',k+1,l+1,m} = \max \Big( T^{(N-1)}_{n,n',k,l,m+1} + T^{(N-1)}_{n,n',k+1,l+1,m}, \nonumber\\ 
 \Xi_N + \Omega_N + \eta_{N} + T^{(N-1)}_{n,n',k,l,m+1} + T^{(N-1)}_{n-1,n'-1,k+1,l+1,m}, \nonumber\\ 
 \Xi_N + \Omega_N + \Omega'_N + 2 \eta_{N} + T^{(N-1)}_{n-1,n'-1,k,l,m+1} + T^{(N-1)}_{n-1,n'-1,k+1,l+1,m} \Big).
\end{eqnarray}
By employing the same reasoning, we obtain
\begin{eqnarray}
  T^{(N)}_{n,n',k+1,l+1,m+1} + T^{(N)}_{n,n',k,l,m} = \max \Big( T^{(N-1)}_{n,n',k+1,l+1,m+1} + T^{(N-1)}_{n,n',k,l,m}, \nonumber\\ 
 \Xi_N + \Omega_N + \Omega'_N + \eta_{N} + T^{(N-1)}_{n-1,n'-1,k+1,l+1,m+1} + T^{(N-1)}_{n,n',k,l,m}, \nonumber\\ 
 \Xi_N + \Omega_N + \Omega'_N + 2 \eta_{N} + T^{(N-1)}_{n-1,n'-1,k+1,l+1,m+1} + T^{(N-1)}_{n-1,n'-1,k,l,m} \Big).
\end{eqnarray}

By the assumption, we have to prove that $T^{(N-1)}_{n,n',k,l,m}$ satisfies the equation:
\begin{eqnarray}\label{mideq1}
 T^{(N-1)}_{n,n',k,l,m+1} + T^{(N-1)}_{n-1,n'-1,k+1,l+1,m} \nonumber\\ 
  = \max \Big( T^{(N-1)}_{n-1,n'-1,k+1,l+1,m+1} + T^{(N-1)}_{n,n',k,l,m} - (R' - \Omega'_N), \nonumber\\
  T^{(N-1)}_{n,n',k,l+1,m} + T^{(N-1)}_{n-1,n'-1,k+1,l,m+1} - (\Omega_N - \Omega'_N), \nonumber\\
  T^{(N-1)}_{n-1,n'-1,k,l+1,m} + T^{(N-1)}_{n,n',k+1,l,m+1} - \Xi_N \Big).
\end{eqnarray}

Here, we consider each case for the values of $\Xi_N$, $\Omega_N$ and $\Omega'_N$. We note that in the case $\Omega'_N = Q_N$, $\Omega_N$ should be equal to $Q_N$ because of $R > R'$. Therefore, we should consider six case $\Xi_N = P_N$ or $S$ and $(\Omega_N, \Omega'_N) = (R, R')$, $(Q_N, R')$ or $(Q_N, Q_N)$. However, in the case $\Xi_N = P_N$ and $\Omega_N = Q_N$, (\ref{ukphier1}) reduces to (\ref{ukpnnpm}), which is already proven. Therefore, it suffices to consider the remaining four cases.

\begin{description}
\item[(I)] The case $\Xi_N = S$ and $(\Omega_N, \Omega'_N) = (Q_N, Q_N)$.
\end{description}
In this case, $T^{(N-1)}_{n,n',k,l,m}$ depends on the form of $n+l+m$ for $n$, $l$, and $m$.  We arrange all of the shifts for the variables $l$ and $m$ to $n'$ and omit these variables. Then, (\ref{mideq1}) is rewritten as
\begin{eqnarray}
 T^{(N-1)}_{n,n'+1,k} + T^{(N-1)}_{n-1,n',k+1} \nonumber\\ 
  = \max \Big( T^{(N-1)}_{n-1,n'+1,k+1} + T^{(N-1)}_{n,n',k} - (R' - Q_N),  T^{(N-1)}_{n,n'+1,k} + T^{(N-1)}_{n-1,n',k+1}, \nonumber\\
  T^{(N-1)}_{n-1,n',k} + T^{(N-1)}_{n,n'+1,k+1} - S \Big)
\end{eqnarray}
Then, we should prove
\begin{eqnarray}
  T^{(N-1)}_{n-1,n'+1,k+1} + T^{(N-1)}_{n,n',k} - T^{(N-1)}_{n,n'+1,k} - T^{(N-1)}_{n-1,n',k+1} \le 0 \\
  T^{(N-1)}_{n-1,n',k} + T^{(N-1)}_{n,n'+1,k+1} - T^{(N-1)}_{n,n'+1,k} - T^{(N-1)}_{n-1,n',k+1} \le S.
\end{eqnarray}
However, these inequalities are satisfied because $T^{(N)}_{n,n',k}$ is a solution of the ultradiscrete KP equation for $(n'; n, l)$.

\begin{description}
\item[(II)] The case $\Xi_N = S$ and $(\Omega_N, \Omega'_N) = (Q_N, R')$.
\end{description}
In this case, $T^{(N-1)}_{n,n',k,l,m}$ depends on $n'+l$ for $n'$ and $l$. Then, (\ref{mideq1}) is rewritten
\begin{eqnarray}
 T^{(N-1)}_{n,n',k,m+1} + T^{(N-1)}_{n-1,n',k+1,m} \nonumber\\ 
  = \max \Big( T^{(N-1)}_{n-1,n',k+1,m+1} + T^{(N-1)}_{n,n',k,m}, \nonumber\\
  T^{(N-1)}_{n,n'+1,k,m} + T^{(N-1)}_{n-1,n'-1,k+1,m+1} - (Q_N - R'), \nonumber\\
  T^{(N-1)}_{n-1,n',k,m} + T^{(N-1)}_{n,n',k+1,m+1} - S \Big).
\end{eqnarray}
Therefore, we should prove
\begin{eqnarray}\label{mideq-a-II-1}
 T^{(N-1)}_{n,n',k,m+1} + T^{(N-1)}_{n-1,n',k+1,m} \nonumber\\ 
  = \max \Big( T^{(N-1)}_{n-1,n',k+1,m+1} + T^{(N-1)}_{n,n',k,m}, \nonumber\\
  T^{(N-1)}_{n-1,n',k,m} + T^{(N-1)}_{n,n',k+1,m+1} - S \Big)
\end{eqnarray}
and
\begin{equation}\label{mideq-a-II-2}
	T^{(N-1)}_{n,n'+1,k,m} + T^{(N-1)}_{n-1,n'-1,k+1,m+1} - T^{(N-1)}_{n,n',k,m+1} - T^{(N-1)}_{n-1,n',k+1,m} \le Q_N - R'.
\end{equation}
Now, (\ref{mideq-a-II-1}) is (\ref{ukphier1}) for $N-1$ and variables $(m; n, k)$ and (\ref{mideq-a-II-2}) is a special case of (\ref{Lem2-1}).

\begin{description}
\item[(III)] The case $\Xi_N = P_N$ and $(\Omega_N, \Omega'_N) = (R, R')$.
\end{description}
$T^{(N-1)}_{n,n',k,l,m}$ depends on $n+k$ for $n$, $k$. Then, (\ref{mideq1}) is rewritten
\begin{eqnarray}
 T^{(N-1)}_{n,n',l,m+1} + T^{(N-1)}_{n,n'-1,l+1,m} \nonumber\\ 
  = \max \Big( T^{(N-1)}_{n,n'-1,l+1,m+1} + T^{(N-1)}_{n,n',l,m}, \nonumber\\
  T^{(N-1)}_{n,n',l+1,m} + T^{(N-1)}_{n,n'-1,l,m+1} - (R - R'), \nonumber\\
  T^{(N-1)}_{n-1,n'-1,l+1,m} + T^{(N-1)}_{n+1,n',l,m+1} - P_N \Big).
\end{eqnarray}
Therefore, we should prove
\begin{eqnarray} \label{mideq-a-III-1}
 T^{(N-1)}_{n,n',l,m+1} + T^{(N-1)}_{n,n'-1,l+1,m} \nonumber\\ 
  = \max \Big( T^{(N-1)}_{n,n'-1,l+1,m+1} + T^{(N-1)}_{n,n',l,m}, \nonumber\\
  T^{(N-1)}_{n,n',l+1,m} + T^{(N-1)}_{n,n'-1,l,m+1} - (R - R') \Big)
\end{eqnarray}
and
\begin{equation}\label{mideq-a-III-2}
  T^{(N-1)}_{n-1,n'-1,l+1,m} + T^{(N-1)}_{n+1,n',l,m+1} - T^{(N-1)}_{n,n',l,m+1} - T^{(N-1)}_{n,n'-1,l+1,m} \le P_N.
\end{equation}
Here, (\ref{mideq-a-III-1}) is a special case of (\ref{ukphier2}) for $N-1$ and (\ref{mideq-a-III-2}) is also a special case of (\ref{Lem2-2}).

\begin{description}
\item[(IV)] The case $\Xi_N = S$ and $(\Omega_N, \Omega'_N) = (R, R')$.
\end{description}
(\ref{mideq1}) is reduced to (\ref{ukphier2}) for $N-1$.

\bigskip
\bigskip

Let us next prove (\ref{ukphier2}). By virtue of a method similar to that in the proof of (\ref{ukphier1}), we should prove
\begin{eqnarray}
  T^{(N-1)}_{n-1,n'-1,k,l,m+1} + T^{(N-1)}_{n-1,n'-1,k+1,l+1,m} = \max \Big( T^{(N-1)}_{n-1,n'-1,k+1,l+1,m+1} + T^{(N-1)}_{n-1,n'-1,k,l,m} - \Omega'_N, \nonumber\\   
  T^{(N-1)}_{n-2,n'-2,k+1,l+1,m+1} + T^{(N-1)}_{n,n',k,l,m} - (P_N - \Xi_N) - (Q_N - \Omega_N), \nonumber\\
  T^{(N-1)}_{n-1,n'-1,k,l+1,m} + T^{(N-1)}_{n-1,n'-1,k+1,l,m+1} -  (R - \Omega_N) + (R' - \Omega'_N), \nonumber\\
  T^{(N-1)}_{n-1,n'-1,k,l+1,m} + T^{(N-1)}_{n-1,n'-1,k+1,l,m+1} -  (S - \Xi_N) \Big).
\end{eqnarray}
Now, by employing Lemma \ref{Lem2}, this equation is reduced to
\begin{eqnarray}
  T^{(N-1)}_{n-1,n'-1,k,l,m+1} + T^{(N-1)}_{n-1,n'-1,k+1,l+1,m} = \max \Big( T^{(N-1)}_{n-1,n'-1,k+1,l+1,m+1} + T^{(N-1)}_{n-1,n'-1,k,l,m} - \Omega'_N, \nonumber\\   
  T^{(N-1)}_{n-1,n'-1,k,l+1,m} + T^{(N-1)}_{n-1,n'-1,k+1,l,m+1} -  (R - \Omega_N) + (R' - \Omega'_N), \nonumber\\
  T^{(N-1)}_{n-1,n'-1,k,l+1,m} + T^{(N-1)}_{n-1,n'-1,k+1,l,m+1} -  (S - \Xi_N) \Big). \label{mideq-b}
\end{eqnarray}
Similar to the proof of (\ref{ukphier1}), we should consider each possible case for the parameters. The case where $\Xi_N = P_N$ and $\Omega_N = Q_N$ is already proven in \cite{Nakata2010}. Therefore, we should prove the remaining four cases.

\begin{description}
\item[(I)] The case $\Xi_N = S$ and $(\Omega_N, \Omega'_N) = (Q_N, Q_N)$.
\end{description}
$T^{(N-1)}_{n,n',k,l,m}$ depends on $n'+l+m$ for $n', l, m$. Now, by virtue of the relation $R > R' > Q_N$, the equation is rewritten as
\begin{eqnarray}
  T^{(N-1)}_{n-1,n',k} + T^{(N-1)}_{n-1,n',k+1} \nonumber\\
	= \max \Big( T^{(N-1)}_{n-1,n'+1,k+1} + T^{(N-1)}_{n-1,n'-1,k} - Q_N, T^{(N-1)}_{n-1,n',k} + T^{(N-1)}_{n-1,n',k+1} \Big).
\end{eqnarray}
Therefore, we should prove
\begin{equation}
  T^{(N-1)}_{n-1,n'+1,k+1} + T^{(N-1)}_{n-1,n'-1,k} - T^{(N-1)}_{n-1,n',k} - T^{(N-1)}_{n-1,n',k+1} \le Q_N. \label{mideq-b-I-1}
\end{equation}
This inequality is obtained by interchanging the roles of the variables between ($n$, $k$) and ($n'$, $m$) in the case $R \ge Q_N$ for (\ref{Lem2-2}).

\begin{description}
\item[(II)] The remaining cases: (\ref{mideq-b}) are reduced to (\ref{ukphier1}) for $N-1$.
\end{description}
\end{Proof}

\section{Generalization to the ultradiscrete KP hierarchy}

Since the dependent variables $k$, $l$ and $m$ do not change in the recursion (\ref{defkphtau}) in the proof in the last section, we can extend the above discussion in two ways. One way is to increase the number of independent variables. By introducing integers $K$ and $L$ satisfying $K, L \ge 1$, and variables $k_1, \ldots, k_K$ and $l_1, \ldots, l_L$, we define $T^{(N)} = T^{(N)}(n, k_1, \ldots, k_K; n', l_1, \ldots l_L)$ as
\begin{eqnarray}
   T^{(N)} = \max \Big( T^{(N-1)}, \eta_N + T^{(N-1)}_{n - 1, n' - 1} \Big) \\
   T^{(0)} \equiv 0
\end{eqnarray}
where
\begin{eqnarray}
  \eta_N = C_N + n P_N + k_1 \Xi_N + \ldots + k_K \Xi_{K, N} + n' Q_N + l_1 \Omega_N + \ldots + \Omega_{L, N}. \label{eta}
\end{eqnarray}
Here, we omit subscripts which do not change in the equation. Applying the proof to the variables ($k_i$; $l_j$, $l_{j'}$) for $j>j'$, we obtain the ultradiscrete KP equation for ($k_i$; $l_j$, $l_{j'}$):
\begin{eqnarray}
  \!\!\!\!\!\!\!\!\!\!\!\!\!\! T^{(N)}_{n, n', k_i, l_j, l_{j'}+1} + T^{(N)}_{n, n', k_i+1, l_j+1, l_{j'}} \nonumber\\
  \!\!\!\!\!\!\! = \max \Big( T^{(N)}_{n, n', k_i+1, l_j+1, l_{j'}+1} + T^{(N)}_{n, n', k_i, l_j, l_{j'}} - R_{j'}, T^{(N)}_{n, n', k_i, l_j+1, l_{j'}} + T^{(N)}_{n, n', k_i+1, l_j, l_{j'}+1} \Big).
\end{eqnarray}
By interchanging the roles of variables $k_i$ and $l_j$, we also obtain the ultradiscrete KP equation for ($k_i$; $l_j$, $l_{j'}$). These equations are ultradiscrete analogues of the basic relation of the discrete KP hierarchy.

The other way is by extending to non-autonomous systems. Let us introduce functions $R_i(x)$ and $S_i(x)$ satisfying $R_i(x) \ge R_j(y)$ and $S_i(x) \ge S_j(y)$ for $\forall x$ ,$\forall y$  and $i > j$ instead of parameters $R_i$ and $S_i$. Then, by denoting $\Xi_{i, N}(k_i) = \min(S_i(k_i), P_N)$ and $\Omega_{i, N}(l_i) = \min(R_i(l_i), Q_N)$, we can replace $k_i \Xi_{i, N}$ and $l_i \Omega_{i, N}$ with $\sum_{k'_i}^{k_i} \Xi_{i, N}(k'_i)$ and $\sum_{l'_i}^{l_i} \Omega_{i, N}(l'_i)$ respectively, in (\ref{eta}). Here $\sum_{x'}^x$ means
\begin{equation}
\sum_{x'}^x = \cases{\sum_{x'=1}^x & for $x>0$ \\ 0 & for $x=0$ \\ -\sum_{x'=x}^{-1} & for $x<0$}.
\end{equation}

Combining these two generalizations yields the following theorem:
\begin{Thm}
The function written as
\begin{eqnarray}
   T^{(N)} = \max \Big( T^{(N-1)}, \eta_N + T^{(N-1)}_{n - 1, n' - 1} \Big) \label{defkphtau2} \\
   T^{(0)} \equiv 0
\end{eqnarray}
for $\eta_N$ given by
\begin{equation}\label{eta}
  \eta_N = C_N + n P_N + \sum_{i=1}^K \sum_{k'_i}^{k_i} \Xi_{i, N}(k'_i) + n' Q_N + \sum_{i=1}^L \sum_{l'_i}^{l_i} \Omega_{i, N}(l'_i),
\end{equation}
solves the non-autonomous ultradiscrete KP hierarchy:
\begin{eqnarray}
  T^{(N)}_{n, n', {\mathfrak k}, {\mathfrak l}, {\mathfrak m}+1} + T^{(N)}_{n, n', {\mathfrak k}+1, {\mathfrak l}+1, {\mathfrak m}} \nonumber\\
  = \max \Big( T^{(N)}_{n, n', {\mathfrak k}+1, {\mathfrak l}+1, {\mathfrak m}+1} + T^{(N)}_{n, n', {\mathfrak k}, {\mathfrak l}, {\mathfrak m}} - {\mathfrak R}'(\mathfrak m), T^{(N)}_{n, n', {\mathfrak k}, {\mathfrak l}+1, {\mathfrak m}} + T^{(N)}_{n, n', {\mathfrak k}+1, {\mathfrak l}, {\mathfrak m}+1} \Big) \label{ukphier3} \\
  T^{(N)}_{n, n', {\mathfrak k}, {\mathfrak l}, {\mathfrak m}+1} + T^{(N)}_{n-1, n'-1, {\mathfrak k}+1, {\mathfrak l}+1, {\mathfrak m}} = \max \Big( T^{(N)}_{n-1, n'-1, {\mathfrak k}+1, {\mathfrak l}+1, {\mathfrak m}+1} + T^{(N)}_{n, n', {\mathfrak k}, {\mathfrak l}, {\mathfrak m}}, \nonumber\\
  T^{(N)}_{n, n', {\mathfrak k}, {\mathfrak l}+1, {\mathfrak m}} + T^{(N)}_{n-1, n'-1, {\mathfrak k}+1, {\mathfrak l}, {\mathfrak m}+1} - {\mathfrak R}(\mathfrak l) + {\mathfrak R}'(\mathfrak m), \nonumber\\ 
  T^{(N)}_{n-1, n'-1, {\mathfrak k}, {\mathfrak l}+1, {\mathfrak m}} + T^{(N)}_{n, n', {\mathfrak k}+1, {\mathfrak l}, {\mathfrak m}+1} - {\mathfrak S}(\mathfrak k) \Big), \label{ukphier4}
\end{eqnarray}
where the tuple of variables (${\mathfrak k}$, ${\mathfrak l}$, ${\mathfrak m}$; ${\mathfrak R}$, ${\mathfrak R}'$, ${\mathfrak S}$) is ($k_i$, $l_j$, $l_{j'}$; $R_j$, $R_{j'}$, $S_i$) or ($l_j$, $k_i$, $k_{i'}$; $S_i$, $S_{i'}$, $R_j$) for $i<i'$ and $j<j'$.
\end{Thm}

\section{Background solutions to the ultradiscrete KP hierarchy}

In this section, we extend the start of the recursion (\ref{defkphtau}). First, we consider the case $P_N < S_i(x)$ and $Q_N < R_j(y)$ for all $i = 1, \ldots, K$, $j=1, \ldots, L$ and $x, y \in \mathbb{Z}$. In this case, $T^{(N)}$ depends only on functions in the form of $n + k_1 + \ldots + k_L$ and $n' + l_1 + \ldots + l_L$. We denote this (two variable) function as $T^0$, i.e,
\begin{equation}\label{defbackground}
	T^{(N)}(n, k_1, \ldots, k_K; n', l_1, \ldots l_L) = T^0(n + k_1 + \ldots + k_K, n' + l_2 + \ldots + l_L).
\end{equation}
Here, by virtue of Lemma \ref{Lem2}, one has
\begin{eqnarray}
	T^0(x+1, y+2) + T^0(x, y) - T^0(x, y+1) - T^0(x+1, y+1) \le Q_N \label{condbg0a} \\
	T^0(x+2, y+1) + T^0(x, y) - T^0(x+1, y) - T^0(x+1, y+1) \le P_N \label{condbg0b}.
\end{eqnarray}
Then, $T^0$ solves the ultradiscrete KP hierarchy (\ref{ukphier3}) and (\ref{ukphier4}) becomes a trivial identity and all necessary inequalities to prove Theorem \ref{mainThm} are generated by the inequalities (\ref{condbg0a}) and (\ref{condbg0b}). We note that we can also append a new soliton parametrized as $P_{N+1} \ge P_N$ and $Q_{N+1} \ge Q_N$ by the recursion (\ref{defkphtau2}). Therefore, we can choose arbitrary $T^0$ satisfying these inequalities for the start of the recursion, i.e., we can choose a $T^0$ that satisfies
\begin{eqnarray}
	T^0(x+1, y+2) + T^0(x, y) - T^0(x, y+1) - T^0(x+1, y+1) \le q \label{condbg1} \\
	T^0(x+2, y+1) + T^0(x, y) - T^0(x+1, y) - T^0(x+1, y+1) \le p \label{condbg2}.
\end{eqnarray}
and $p \le P_1$ and $q \le Q_1$. However, it should be noted that we do not need to restrict $p$ and $q$ to $p, q >0$, i.e., we can construct solutions even if the suprema of the left hand sides of (\ref{condbg1}) and (\ref{condbg2}) are negative. This solution is called a ``background", because all solitons evolve over it. Different from soliton solutions,  backgrounds can keep their structure under addition because the left hand sides of (\ref{condbg1}) and (\ref{condbg2}) are linear for $T^0$. Now, note that the solutions of the ultradiscrete KP equation in \cite{Nakata2010} are special cases of the solution defined in (\ref{defkphtau}). The backgrounds are further special cases of the solutions in \cite{Nakata2010}.

\section{Application: Box and Ball systems with time-dependent Carrier}

It is known that the standard BBS has several degrees of freedom, for example, the capacity of balls for boxes at each site and for carriers at each time. By setting reductions properly, we can obtain exact solutions of this system with backgrounds. 

We consider the case $K=1$, $M=3$ and assume that $T^{(N)}$ depends on $n+n'+k_1+l_1$ for $n$, $n'$, $k_1$ and $l_1$, i.e., we consider the case where $R_1( l_1)$ and $S_1(k_1)$ are sufficiently large. We now introduce new independent variables $n+n'+k_1+l_1=x$, $l_2=t$ and $l_3=-j$ and new dependent variables $T^{(N)}_{x, l_2, l_3} = F^t_j$, $T^{(N)}_{x+1, l_2, l_3} = G^t_j$, $R_2(l_2) = A^t$ and $R_3(l_3) = B_j$. Note that because of the constraints on $R_j(x)$,  $R_j(y)$ of section 3, we have that $A^t \ge B_j$, which means that the capacity of the carriers is greater than that of the boxes. Then, the non-autonomous ultradiscrete KP equation for ($k_1$; $l_1$, $l_2$) and ($k_1$; $l_1$, $l_3$) is rewritten as
\begin{eqnarray}
  F^{t+1}_{j+1} + G^t_j = \max \Big( F^t_{j+1}  +G^{t+1}_j - A^t, F^{t+1}_j + G^t_{j+1} \Big) \label{backlund1} \\
  F^t_j + G^{t+1}_{j+1} = \max \Big( F^t_{j+1}  +G^{t+1}_j - B_j, F^{t+1}_j + G^t_{j+1} \Big) \label{backlund2}.
\end{eqnarray}
Here, we denote $U^t_j = F^t_{j+1} + G^t_j - F^t_j - G^t_{j+1}$ and set the boundary condition: $U^t_j = 0$ for $|j| \gg 1$. Due to these equations,
\begin{eqnarray}
  -G^{t+1}_j - F^{t+1}_j + \max \Big( F^t_{j+1}  +G^{t+1}_j - A^t, F^{t+1}_j + G^t_{j+1} \Big) \nonumber\\
  \quad - \Big( G^t_j - F^t_j - G^t_{j+1} - G^{t+1}_j + \max \Big( F^t_{j+1}  +G^{t+1}_j - B_j, F^{t+1}_j + G^t_{j+1}\Big) \Big)
\end{eqnarray}
equals $U^{t+1}_j$. However, by virtue of the boundary condition, we also express this as 
\begin{eqnarray}
  \!\!\!\!\!\!\!\!\!\!\!\!\!\!\!\! \min \left( B_j - U^t_j, \sum_{n=-\infty}^{j-1} U^t_n - \sum_{n=-\infty}^{j-1} U^{t+1}_n \right) + \max \left( \sum_{n=-\infty}^j U^t_n - \sum_{n=-\infty}^{j-1} U^{t+1}_n - A^t, 0 \right).
\end{eqnarray}
Therefore, one has
\begin{eqnarray}\label{BBSC}
  \!\!\!\!\!\!\!\!\!\!\!\!\!\!\!\! U^{t+1}_j = \min \left( B_j - U^t_j, \sum_{n=-\infty}^{j-1} U^t_n - \sum_{n=-\infty}^{j-1} U^{t+1}_n \right) + \max \left( \sum_{n=-\infty}^j U^t_n - \sum_{n=-\infty}^{j-1} U^{t+1}_n - A^t, 0 \right), \nonumber\\
\end{eqnarray}
which is the time evolution rule of the Box and Ball system with Carrier (BBSC). Solutions to this equation are obtained by setting parameters $P_i = Q_i$ for $S_1(x) \gg 1$. Soliton solutions for this system were first presented in \cite{TakahashiMatsukidaira} and the direct relationship to the discrete case is shown in \cite{IsojimaKuboMurataSatsuma} by ultradiscretizing soliton solutions of the discrete modified KdV equation. We note that we can also construct solutions of the BBSC multi-kinds of boxes and balls by properly setting different reduction conditions.

Due to the specialization of parameters $P_N$ and $Q_N$, the background solution satisfying (\ref{condbg1}) and (\ref{condbg2}) is expressed as
\begin{eqnarray}
	T^0(n, n') = \sum_{\tilde{n} \in \mathbb{Z}} U_{j'} \max( n+n' - j', 0),
\end{eqnarray}
where $U_{\tilde{j}}$ is a function satisfying $U_{\tilde{j}} + U_{\tilde{j}+1} \le 0$.
\begin{figure}
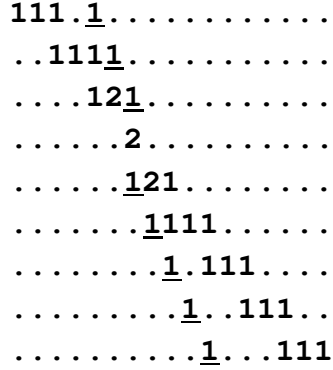

\begin{center}
\begin{tabular}{c}
{\usefont{OT1}{pcr}{b}{n}111.\underline{1}............} \\
{\usefont{OT1}{pcr}{b}{n}..111\underline{1}...........} \\
{\usefont{OT1}{pcr}{b}{n}....12\underline{1}..........} \\
{\usefont{OT1}{pcr}{b}{n}......2..........} \\
{\usefont{OT1}{pcr}{b}{n}......\underline{1}21........} \\
{\usefont{OT1}{pcr}{b}{n}.......\underline{1}111......} \\
{\usefont{OT1}{pcr}{b}{n}........\underline{1}.111....} \\
{\usefont{OT1}{pcr}{b}{n}.........\underline{1}..111..} \\
{\usefont{OT1}{pcr}{b}{n}..........\underline{1}...111}
\end{tabular}
\caption{An example of a time evolution of the BBSC with backgrounds. Dots denote $0$ and underscores negative values.}
\label{pic1}
\end{center}
\end{figure}
Figure \ref{pic1} depicts a solution parametrized by $U_{\tilde{n}} = -\delta_{\tilde{n}, 0}$, $P_1 = 3$, $C_1 = 0$, $B_j \equiv 1$ and $A^t \equiv 2$.

\section{Compatibility condition of the non-autonomous ultradiscrete KdV equation}

It is known that the reduction procedure to obtain the discrete modified KdV equation \cite{KakeiNimmoWillox} is the same as that to obtain the non-autonomous discrete KdV equation \cite{Matuura, KajiwaraOhta}. We expect that the BBSC and an ultradiscretization of the non-autonomous discrete KdV equation are equivalent at the level of the $\tau$-function. Since the discrete KdV equation is obtained by the compatibility condition of the discrete modified KdV equation, it is expected that (\ref{backlund1}) and (\ref{backlund2}) are an ultradiscrete analogue of Lax form for ultradiscretization of the non-autonomous KdV equation and that the compatibility condition yields this equation. We note that such a compatibility condition of ultradiscrete systems is already discussed in \cite{ShinzawaHirota}.

\bigskip

By setting $\Phi^t_j = F^t_j - G^t_j$ and $V^t_j = G^{t+1}_j + G^t_{j+1} - G^{t+1}_{j+1} - G^t_j$, (\ref{backlund1}) and (\ref{backlund2}) are rewritten as
\begin{eqnarray}
  \Phi^t_j - V^t_j = \max \Big( \Phi^t_{j+1} - A^t, \Phi^{t+1}_j \Big) \label{ism2} \\
  \Phi^{t+1}_{j+1} - V^t_j = \max  \Big( \Phi^t_{j+1} - B_j, \Phi^{t+1}_j \Big). \label{ism1}
\end{eqnarray}
However, we cannot eliminate $\Phi^t_j$ from these equations. To avoid this problem, we employ (\ref{ukphier2}) for $S \gg 1$:
\begin{eqnarray}
  \Phi^t_j - V^t_j = \max \Big( \Phi^{t+1}_{j+1} - V^t_j - A^t + B_j, \Phi^{t+1}_j \Big) \label{ism3}
\end{eqnarray}
instead of (\ref{ism2}). Since we can recover (\ref{ism2}) by substituting (\ref{ism1}) to (\ref{ism3}), we start by considering the compatibility condition from these equations. Here, taking the maximum of $\Phi^t_{j+1} - B_j$ with each side hand of (\ref{ism3}), we obtain
\begin{eqnarray}
	\max \Big( \Phi^t_{j+1} - B_j,  \Phi^t_j - V^t_j  \Big) &= \max \Big( \Phi^{t+1}_{j+1} - V^t_j - A^t + B_j, \Phi^{t+1}_j, \Psi^t_{j+1} - B_j \Big) \nonumber\\
								    &= \max \Big( \Phi^{t+1}_{j+1} - V^t_j - A^t + B_j,  \Phi^{t+1}_{j+1} - V^t_j \Big) \nonumber\\
								    &= \Phi^{t+1}_{j+1} - V^t_j \label{ism4}
\end{eqnarray}
because of (\ref{ism1}) and $A^t \ge B_j$. We now consider the compatibility condition for (\ref{ism3}) and (\ref{ism4}). By substituting (\ref{ism3}) to (\ref{ism4}), we obtain
\begin{eqnarray}
	\Phi^{t+1}_{j+1} - V^t_j = \max \Big( \Phi^{t+1}_{j+1} - V^t_j - A^t + B_j, \Phi^t_{j+1}, \nonumber\\
					\qquad\qquad\qquad V^{t+1}_j - B_j + \max \Big( \Phi^{t+2}_{j+1} - V^{t+1}_j - R_{1, t+1} + B_j, \Phi^{t+1}_{j+1} \Big) \Big) \nonumber\\
					= \max \Big( \max \Big( -V^t_j - A^t + B_j, V^{t+1}_j - B_j \Big) + \Phi^{t+1}_{j+1}, \Phi^t_{j+1}, \Phi^{t+2}_{j+1} - A^{t+1} \Big).
\end{eqnarray}
By substituting (\ref{ism4}) to (\ref{ism3}) for $t \to t+1$ and $j \to j+1$, we obtain
\begin{eqnarray}
	\Phi^{t+1}_{j+1} - V^{t+1}_{j+1} = \max \Big( - A^{t+1} + B_{j+1} + \max \Big( \Phi^{t+1}_{j+1} - V^{t+1}_{j+1}, \Phi^{t+2}_{j+1} - B_{j+1} \Big), \nonumber\\
		\qquad\qquad\qquad\qquad  V^t_{j+1} + \max \Big( \Phi^t_{j+1} - V^t_{j+1}, \Phi^{t+1}_{j+1} - B_{j+1}\Big) \Big) \nonumber\\
						 = \max \Big( \max \Big( - A^{t+1} + B_{j+1} - V^{t+1}_{j+1}, V^t_{j+1} - B_{j+1} \Big) + \Phi^{t+1}_{j+1}, \nonumber\\ 
		\qquad\qquad\qquad - A^{t+1} + \Phi^{t+2}_{j+1}, \Phi^t_{j+1} \Big).
\end{eqnarray}
Therefore, $\max \Big( \max \Big( -V^t_j - A^t + B_j, V^{t+1}_j - B_j \Big) + \Phi^{t+1}_{j+1},  \max \Big( - A^{t+1} + B_{j+1} - V^{t+1}_{j+1}, V^t_{j+1} - B_{j+1} \Big) + \Phi^{t+1}_{j+1}, \Phi^t_{j+1}, \Phi^{t+2}_{j+1} - A^{t+1} \Big)$ can be expressed by two ways:
\begin{eqnarray}
	\max \Big( \Phi^{t+1}_{j+1} - V^{t+1}_{j+1}, \max \Big( -V^t_j - A^t + B_j, V^{t+1}_j - B_j \Big) + \Phi^{t+1}_{j+1} \Big) \nonumber\\
		= \max \Big( \Phi^{t+1}_{j+1} - V^t_j, \max \Big( - A^{t+1} + B_{j+1} - V^{t+1}_{j+1}, V^t_{j+1} - B_{j+1} \Big) + \Phi^{t+1}_{j+1} \Big)
\end{eqnarray}
By subtracting $\Phi^{t+1}_{j+1}$ from each hand side, we obtain
\begin{eqnarray}
  \max \Big( - V^{t+1}_{j+1}, - V^t_j - A^t + B_j, V^{t+1}_j - B_j \Big) \nonumber\\
	 = \max \Big( - V^{t+1}_{j+1} - A^{t+1} + B_{j+1},  V^t_{j+1} - B_{j+1}, - V^t_j \Big), \label{nonlinearukdv1}
\end{eqnarray}
which is the compatibility condition for equations (\ref{ism3}) and (\ref{ism4}). However, this equation is different from that obtained by naively ultradiscretizing the non-autonomous discrete KdV equation in \cite{KajiwaraOhta}. Because of $A^t > B_j$ for $\forall t$ and $\forall j$, (\ref{nonlinearukdv1}) is simplified to
\begin{eqnarray}\label{ukdvnonlinear}
  \max \Big( - V^{t+1}_{j+1}, V^{t+1}_j - B_j \Big) = \max \Big( V^t_{j+1} - B_{j+1}, - V^t_j \Big),
\end{eqnarray}
which looks better as an ultradiscretization but the contribution of $A^t$ has disappeared. We also note that (\ref{nonlinearukdv1}) and (\ref{ukdvnonlinear}) are not evolution equations.

\begin{figure}
\begin{center}
\begin{tabular}{c c}
\begin{minipage}{0.5\hsize}
\begin{center}
\begin{tabular}{c}
{\usefont{OT1}{pcr}{b}{n}1....1.........} \\
{\usefont{OT1}{pcr}{b}{n}.1....1........} \\
{\usefont{OT1}{pcr}{b}{n}..1....1.......} \\
{\usefont{OT1}{pcr}{b}{n}...11...1......} \\
{\usefont{OT1}{pcr}{b}{n}.....11..1.....} \\
{\usefont{OT1}{pcr}{b}{n}.......11.1....} \\
{\usefont{OT1}{pcr}{b}{n}.........1.11..} \\
{\usefont{OT1}{pcr}{b}{n}..........1..11}
\end{tabular}
\caption{Time evolution of $V^t_j$.}
\label{pic2}
\end{center}
\end{minipage} & \begin{minipage}{0.5\hsize}
\begin{center}
\begin{tabular}{c}
{\usefont{OT1}{pcr}{b}{n}1111.11........} \\
{\usefont{OT1}{pcr}{b}{n}.1111.11.......} \\
{\usefont{OT1}{pcr}{b}{n}..1111.11......} \\
{\usefont{OT1}{pcr}{b}{n}...1111.11.....} \\
{\usefont{OT1}{pcr}{b}{n}.....111.111...} \\
{\usefont{OT1}{pcr}{b}{n}.......11.1111.} \\
{\usefont{OT1}{pcr}{b}{n}.........1.1111} \\
{\usefont{OT1}{pcr}{b}{n}..........1..11} \\
{\usefont{OT1}{pcr}{b}{n}...........1...}
\end{tabular}
\caption{Time evolution of $U^t_j$}
\label{pic3}
\end{center}
\end{minipage}
\end{tabular}
\end{center}
\end{figure}

\bigskip

We next observe the behavior of $V^t_j$ numerically by setting parameters $N=2$, $P_1 = 1$, $P_2 = 5$, $C_1 = 0$, $C_2 = -3$, $B_j \equiv 1$, $A^t \equiv 1$ for $t < 0$, $A^t \equiv 2$ for $t \ge 0$. Figure \ref{pic2} depicts $V^t_j$ in the range $-3 \le t \le 4$. Two blocks of balls travel with at the same speed $1$ when $t<0$. But when $t=0$, a ball is injected the former block and passes the latter one finally. This phenomenon can be explained using the ultradiscrete Miura transformation of \cite{KuboIsojimaMurataSatsuma}.
\begin{Thm}[Ultradiscrete Miura transformation]
The dependent variables $U^t_j$ and $V^t_j$ satisfy the relationship:
\begin{equation}\label{udmiura}
	V^t_j = \min \Big( U^{t+1}_j, B_j - U^t_j \Big).
\end{equation}
\end{Thm}

The following ultradiscrete-closed proof is originally presented in \cite{Kubo}.
\begin{Proof}
By subtracting left hand side from right hand side, we obtain
\begin{equation}
	\min \Big( F^{t+1}_{j+1} + G^t_j - F^{t+1}_j - G^t_{j+1}, F^t_j + G^{t+1}_{j+1} - F^t_{j+1} - G^{t+1}_j + B_j \Big).
\end{equation}
Here, by employing (\ref{backlund1}) and (\ref{backlund2}), one can rewrite
\begin{eqnarray}
	\!\!\!\!\!\!\!\!\!\!\!\!\!\!\!\!\!\!\!\!\!\!\!\!\!\!\! \min \Big( \max \Big( F^t_{j+1} + G^{t+1}_j - F^{t+1}_j - G^t_{j+1} - A^t, 0 \Big), \max \Big( 0, F^{t+1}_j + G^t_{j+1} - F^t_{j+1} - G^{t+1}_j + B_j \Big) \Big), \nonumber\\
\end{eqnarray}
which is identically equal to $0$ because of $A^t \ge B_j$.
\end{Proof}

Due to the ultradiscrete Miura transformation (\ref{udmiura}),  the case where $V^t_j = 1$ is achieved only if $U^{t+1}_j = 1$ and $U^t_j = 0$. Therefore, the length of a block increases at $t=0$ because of an increment of capacity of the carrier and the velocity of blocks in the BBSC (\ref{BBSC}). Figure \ref{pic3} depicts the time evolution of $U^t_j$ which is obtained by (\ref{udmiura}) from $V^t_j$ in Figure \ref{pic2}. The contribution of $A^t$ is hidden in the degrees of freedom of solutions to (\ref{ukdvnonlinear}).

To end this section, we stress that $V^t_j$ does {\it not} satisfy the ultradiscrete KdV equation known as the time evolution rule for the standard BBS:
\begin{equation}\label{standardBBS}
	V^{t+1}_j = \min \left( B_j - V^t_j, \sum_{j'=-\infty}^{j-1} \Big(V^t_{j'} - V^{t+1}_{j'} \Big) \right)
\end{equation}
because this equation is obtained from by taking reductions $T_{n, n', k, l, m} = F^{n+n'+k+l}_m =: F^t_j$ (and $V^t_j$ is the same) from the ultradiscrete KP equation for ($n$; $n'$, $m$). However, any $V^t_j$ that satisfies (\ref{standardBBS}) also solves (\ref{ukdvnonlinear}), because one can represent $\min \Big( B_j - V^{t+1}_j, B_{j+1} - V^t_{j+1}, \sum_{j'=-\infty}^j \big(V^t_{j'} - V^{t+1}_{j'} \big) \Big)$ in two ways.

\section{Concluding Remarks}

In this paper, we proposed a recursive representation of solutions to an ultradiscrete analogue of the discrete KP hierarchy and its background solutions as a starting point for the recursion.

For the equation, we took two different types of independent variables $k_i$ and $l_j$ and chose one variable from one type and two variables from the other. In the proof, we employed the properties of two variables which belong to the same type. However, solutions no longer solve the equation if we choose all three variables from one type. The difference of type plays an important role in the ultradiscrete equations.

Equation (\ref{ukphier2}) is considered to be obtained by ultradiscretizing a combination of some equations in the discrete KP hierarchy. In fact, we employ the reduction of (\ref{ukphier2}) as a Lax form for the ultradiscrete KdV equation in section 6. Different from the discrete equations, ultradiscrete equations are not easily combined because of the $\max$ operators. Employing this reduction approach, we can obtain variations of ultradiscrete soliton equations which have been previously been obtained from discrete equations by an ultradiscrete, closed, approach.

As discussed in section 3, we can consider that the recursion (\ref{defkphtau}) does not depend on any other independent variables except $n$ and $n'$. In this case however, equation (\ref{ukphier1}) is essentially not a difference equation but just an ordinary algebraic equation. Since the number of variables is greater than that of the equations, its solutions form a 1-parameter family. If this parameter satisfies a relation corresponding to (\ref{ukphier2}), we can obtain new solutions by applying the recursion to these solutions. We believe that this can be explained by some elementary approach, for example, a geometrical one, which should represent the essence of the ultradiscrete soliton equations.

\section*{Acknowledgment}
The author would like to thanks  Professors T. Tokihiro and R. Willox for helpful comments. This work was supported by Platform for Dynamic Approaches to Living System from the Ministry of Education, Culture, Sports, Science and Technology, Japan.

\section*{References}

\end{document}